\documentclass[journal]{IEEEtran}

\newlength{\figwidth} 
\usepackage[utf8x]{inputenc}
\usepackage[english]{babel}
\usepackage{verbatim} 
\usepackage{varioref}
\usepackage{psfrag}
\usepackage{epsf}
\usepackage{graphics}
\usepackage{graphicx}
\usepackage{amsbsy}
\usepackage{amssymb}
\usepackage{amscd}
\usepackage{amsmath}
\usepackage{amsthm}
\usepackage{enumerate} 
\usepackage{color}
\usepackage{epstopdf}
\usepackage{threeparttable}

\usepackage{cite}
\usepackage{url}

\def\mN{{{\mathcal{N}}}}

\title{Suburban Fixed Wireless Access Channel Measurements and Models at 28 GHz for 90\% Outdoor Coverage}

\author{ 
Jinfeng Du, \emph{Member, IEEE}, Dmitry Chizhik, \emph{Fellow, IEEE}, Rodolfo Feick, \emph{Senior Member, IEEE},\\  
Mauricio Rodr\'iguez, \emph{Senior Member, IEEE}, Guillermo Castro, and Reinaldo A. Valenzuela, \emph{Fellow, IEEE}
\thanks{Jinfeng Du, Dmitry Chizhik and Reinaldo. A. Valenzuela are with Nokia Bell Labs, Holmdel, NJ 07733, USA (e-mail: 
\{jinfeng.du, dmitry.chizhik, reinaldo.valenzuela\}@nokia-bell-labs.com).}
\thanks{Rodolfo Feick is with Department of Electronics, Universidad T\'ecnica Federico Santa Maria, Valpara\'iso, Chile (e-mail: rodolfo.feick@usm.cl).}
\thanks{Mauricio Rodr\'iguez and Guillermo Castro are with Escuela de Ingenier\'ia El\'ectrica de la Pontificia Universidad Cat\'olica de Valpara\'iso, Valpara\'iso, Chile (e-mail: \{mauricio.rodriguez.g, guillermo.castro\}@pucv.cl).} 
\thanks{This work has been presented in part at 2018 IEEE Symposium on Antennas and Propagation \cite{TAP_23}.}
} 

\begin{document}

\maketitle

\begin{abstract} 
Achieving adequate coverage with high gain antennas is key to realizing the full promise of the wide bandwidth available at {cm/mm} bands. We report extensive outdoor measurements at 28 GHz in suburban residential areas in New Jersey and Chile, with over 2000 links measured for same-street link {types} (vegetation blocked LOS) from 13 streets and other-street link {types} (true NLOS) from 7 streets, using a specialized narrowband channel sounder at ranges reaching 200 m. The measurements, applicable to fixed wireless access, involved a 55$^\circ$ transmit antenna placed on the exterior of a street-facing window and a 10$^\circ$ receive horn antenna spinning on top of a van mast at 3 m height, emulating a lamppost-mounted base station. Measured path gain-distance dependence is well represented by power-law models, and azimuth gains at the base are degraded through scattering by more than 4.3 dB for 10\% of links. It was found that, with 51 dBm EIRP at the base station and 11 dBi antenna at an outdoor mounted terminal, 1 Gbps downlink rate can be delivered up to 100 m from a base station deployed in the same street with 90\% coverage guarantee.
\end{abstract}
 

\section{Introduction}\label{sec:introduction}

{Newly allocated vast spectrum at cm/mm bands offers the opportunity for very high data rates. }
 However, higher free-space, scattering, transmission and diffraction losses in these bands {imposed significant challenge for next generation wireless systems} \cite{TAP_19}. To overcome such losses one cannot count on the nominal (free-space) gain of an antenna, as this may be significantly reduced by scattering \cite{TAP_8, TAP_14}. Moreover, temporal fluctuation can pose beam adaptation requirements. 
{Therefore, propagation losses at these frequencies must be accurately understood and the effectiveness of antenna directivity/gain to compensate these losses must be experimentally validated to support proper system design to meet the quality of service requirements for end-users.
 We focus on one of the early 5G candidate deployment scenarios for Fixed Wireless Access (FWA) in a suburban setting where a street-mounted base station (BS) at lamppost height} provides high-speed broadband service to user terminals mounted on the exterior of single-family homes along a street, in the presence of vegetation. 
We collected a statistically significant set of path gains, effective antenna gains as well as {temporal variation of received power} to reliably describe performance at 90\% of locations in a typical outdoor suburban environment.

Numerous measurement campaigns have largely concentrated on urban areas, particularly at higher frequencies \cite{TAP_1,TAP_2,TAP_3,TAP_4,TAP_5,TAP_6,TAP_15,TAP_16,TAP_17,TAP_18,TAP_19}. A more comprehensive survey of measurements at mmWave bands can be found in \cite{TAP_1,TAP_16,TAP_20}. The data is generally separated into line of sight (LOS) and non-line of sight (NLOS) categories. Path loss for each {one} was characterized with slope-intercept models.  Path loss models presented in \cite{TAP_5} were based on measurements in urban canyons in multiple cities. Same street LOS measurements were unobstructed by vegetation, with NLOS defined as being around-the-corner, and the 3GPP 36.814 UMi NLOS model with Manhattan grid layout was modified to include a frequency dependent turn loss at the corner. This approach was also adopted in \cite{TAP_15,TAP_16,TAP_17}, where the around corner model was extended to many corners based on ray-tracing data from a 3D building database (cars, street signs, and billboards are not considered). 

The majority of the previous outdoor measurements at cm/mm bands were carried out in urban city street canyons (tall and continuous buildings on both side, almost no vegetation blockage) or university/industry campus (open space), which is quite different from suburban residential areas where {the} blockage of trees/bushes is an important factor for signal attenuation. In \cite{TAP_18} suburban measurements were carried out by placing a transmitter at a height of 7.5 m in a street (mimicking a base), and moving the receiver either along the same street, or into two nearby streets that are perpendicular to the street with the transmitter. The slope for same-street path loss was found to be close to 3 when using an omni-antenna, and higher path loss was reported when using a directional antenna (peak power in angular domain was used).  

\subsection{Objectives and scope of the work}

 {The objective is to collect a massive data base  to reliably describe the performance at 90\% of locations in suburban FWA settings.} We used a specially constructed 28 GHz narrowband channel sounder, allowing rapid measurements of directional channel response, with large dynamic range. Fast data collection allows both an assessment of channel {temporal variation} as well as permitting rapid gathering of data. We separate, as is traditional, the path gain measurements from the directional aspects of the channel. This is done by averaging over all directions the received power to provide an effective average power that an omni antenna would have measured. Directional (azimuthal) gain is estimated separately from normalized angular spectra. We measured more than 2,000 links in two towns (one in NJ and the other in Chile),  {including a total of over 14 million individual power measurements.} They consist of same-street links (vegetation blocked LOS) from 13 streets, and other-street links (true NLOS) from 7 streets.  {The main contribution of our work includes the formulation propagation models for path gains, effective antenna gains as well as temporal variation of received power. To validate the statistical significance of our results we collected enough data to yield confidence intervals that guarantee that the values reported for the model parameters are not the result of purely anecdotal conditions. We also verified that model parameters obtained from subsets of our data were consistent with each other. For example,} despite the rich diversity of house/tree types and density, the power law slope-intercept fits to measured path loss from NJ and from Chile are similar.

{The measurement dataset and path loss models reported in this work can be very useful to both academic and industrial researchers/engineers working on FWA related deployment and applications. It improves existing databases with a significant amount of new measured links and provides additional insights for channel modeling in suburban FWA scenarios.}

The rest of the paper is organized as follows: A brief description of the measurement equipment and environment is presented in Sec.~\ref{sec:measurement}, and path gain measurement and models for suburban street canyon are presented in Sec.~\ref{sec:pathGain}. Effective azimuth antenna gain results are reported in Sec.~\ref{sec:azim} and the benefit of re-aiming after each azimuth scan is quantified in Sec.~\ref{sec:switch}.  Outdoor-to-outdoor downlink data rates with service guarantee for 90\% of locations is evaluated in Sec.~\ref{sec:Rate} and conclusions are presented in Sec.~\ref{sec:conclusion}.

\section{Measurement equipment and environment}\label{sec:measurement}  

\subsection{Measurement equipment}\label{sec:setup} 

  To maximize data collection speed and link budget, we constructed a narrowband sounder, transmitting a 28 GHz  continuous-wave (CW)  tone at 22 dBm power. The transmitter {(Tx)} has a 10 dBi horn with 55$^\circ$ half-power beamwidth in both elevation and azimuth. The receiver {(Rx)} has a {24 dBi (10$^\circ$)} horn, connected to a low-noise amplifier, a mixer, and a power meter, with a 20 kHz receive bandwidth and effective noise figure of 5 dB{, which leads to a noise floor of -126 dBm}.  The {Tx and Rx have separate, internally referenced, phase-locked dielectric resonator oscillators (PLDROs),} with a frequency accuracy better than $10^{-7}$. {At the 28 GHz frequency this means a worst-case drift of $\pm 2.8$ kHz, well within the IF filter’s bandwidth, as was repeatedly verified in calibration measurements with a spectrum analyzer.} The Rx was mounted on a rotating platform {spinning at up to 300 rpm}, allowing a full angular scan every 200 ms. Power samples {are recorded}  at a rate of 740 samples/second, using an onboard computer, {which also records angular position with a resolution of 1$^\circ$. At full speed this results in one power sample every 2.5$^\circ$, which is adequate given the 10$^\circ$ beamwidth of the receiver. 	}
	
	The system was repeatedly calibrated in the lab and anechoic chamber to assure measurement power accuracy. {Our power meter has a 0.15 dB resolution and during short-term measurements this was indeed observed to be the residual power fluctuation. The long-term measurement uncertainty due to the combined effect of transmit power and receiver gain drift did not exceed $\pm 1$ dB as determined by comparing multiple calibration measurements during the data collection process.} 
	The full dynamic range of the {Rx} (from noise floor to 1 dB compression point) was found to be 50 dB, extensible to 75 dB using switchable amplifiers. 
	
	{Considering a minimum 10 dB measurement SNR, 22 dBm transmit power and -126 dBm noise floor would permit a reliable measurement of path loss} up to 138 dB (200 m range with 30 dB excess loss) {using non-directional antennas. By combining} removable Tx attenuators {with switchable Rx amplifiers,}  measurable path loss  ranged from  61 dB (1 m in free space) to  172 dB {considering the nominal gain of the antennas we used. Note that such high link budget comes at the cost that the narrowband sounder cannot resolve multiple paths in time. Distance was measured using tape, laser rangefinder and GPS to assure errors no larger than 2\%.} 
	 
\begin{figure} 
	\centering
		\includegraphics[width=0.9\figwidth]{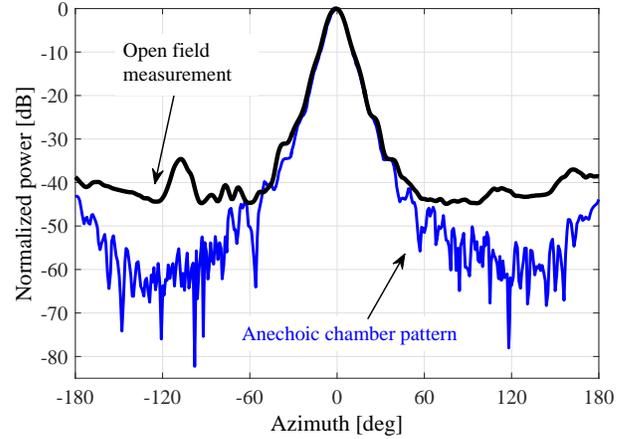}
	\caption{LOS measurement in open field matches the anechoic measurement to -40 dB.}
	\label{fig:filed_LOS}
\end{figure}

	The {Rx} horn was tested in an open field at a range of 40 meters from the {Tx}. The measured receive pattern is seen in Fig.~\ref{fig:filed_LOS} to be within 1 dB of that measured in the anechoic chamber, down to -40 dB.
	
 \subsection{Measurement environment}\label{sec:scenario}

\begin{figure} 
	\centering
		\includegraphics[width=0.9\figwidth]{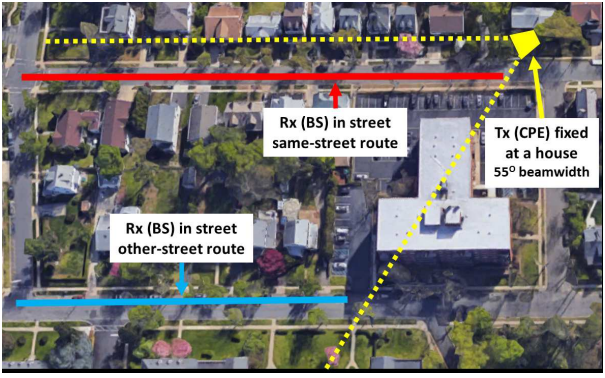} 
	\caption{Top view of a typical measurement area in a NJ suburb.  {Rotary} Rx (BS) was mounted 3 m above ground, moving along the street (red line for same-street and blue line for ``other-street'' measurement), range 20 to 200 m. Tx (CPE) was fixed near house, {aiming at a fixed angle illuminating the entire Rx route with its 55$^\circ$-wide beam (dotted yellow lines).}}
	\label{fig:street}
\end{figure}

 Measurements were performed at 28 GHz in suburban residential areas in New Jersey (NJ) and Vi\~na del Mar, Chile. The NJ suburb has mostly 2-story wood-frame homes along 25 m-wide streets. The vegetation along the streets and on properties consisted of tall trees and bushes, representative of North-Eastern US. The Chilean suburbs have narrower streets and denser houses as compared to NJ, and the front yards are often surrounded by a metal fence and low-height bushes.  

The {Rx} was placed on a van mast 3 m above ground, emulating a lamppost mounted {base station (BS).  The Tx, emulating a customer premises equipment (CPE)}, was placed near the exterior of a street-facing window aiming at {a fixed angle along the street,} illuminating the {entire range of positions successively occupied by} the Rx, {to assure the Rx was always within the 55$^\circ$-wide Tx beam}. A typical measurement geometry is illustrated in Fig.~\ref{fig:street}, where the ``same-street'' route {refers to the scenario when} the serving BS and the CPE are on the same street, and the ``other-street'' scenario is when the BS and the CPE are on a pair of contiguous parallel streets. Measurements were collected every 1 m {(3 m at distances beyond 70 m)} for ranges from 20 m to 200 m. 
{At each Rx location} a 10-second record containing narrowband receive power as a function of azimuth angle and time was obtained, consisting of at least 37 full azimuthal scans over 360 degrees {with one-degree resolution}.   A total of over 1700 {such} link measurements were made, using 6 houses in NJ and 7 houses in Chile, for the ``same-street'' deployment scenario. {Despite both terminals being} on the same street,  {the presence of vegetation implies that some links are LOS and others NLOS, a condition we recorded for each measurement as will be described later.}  The ``other-street'' scenario was measured using 4 streets in NJ and 3 streets in Chile {for a total of over 180 links. In these cases all the links were NLOS due to the combined effect of vegetation and houses.}

\subsection{Computation of path gain and effective azimuthal gain from measurements}

The {rotary Rx} horn {offers a set of power} measurements as a function of azimuth and time. Our goal is to determine overall pathloss suffered by the signal, and, separately the distribution of the signal over {the azimuth} angle, to determine beamforming effectiveness.
It can be shown \cite{TAP_8} that averaging power measurements $P(\phi)$  over azimuth $\phi$ using a directive antenna has the same expected value as local spatial average ${<}P_{\text{omni}}{>}$  of an azimuthally omnidirectional antenna:
\begin{align}
{<}P(\phi){>} \triangleq  \frac{1}{2\pi}\int_0^{2\pi} P(\phi)d\phi={<}P_{\text{omni}}{>}.
\end{align} 
This remains so despite the beam overlap between successive azimuthal aiming of the rotating horn antenna~\cite{TAP_8}.

Path gain $P_{\text{G}}$  is calculated by averaging the received power over all azimuthal angles. By removing the transmitted power $P_{\text{T}}$, {the transmit antenna gain $G_{\text{T}}$}, and the elevation gain $G_{\text{elev}}$, we obtain
\begin{align}
P_{\text{G}} = {<}P(\phi){>} - P_{\text{T}} - {G_{\text{T}}} - G_{\text{elev}}, \ [\text{dB}].
\label{eqn:PL}
\end{align}
The elevation gain is obtained from the total antenna gain $G_{\text{tot}}$ (as measured in the anechoic chamber):
\begin{align}
G_{\text{elev}} = G_{\text{tot}} - G_{\text{azim}},\ [\text{dB}],
\end{align}
where the {nominal azimuthal antenna gain} is given by the peak-to-average ratio of the antenna pattern $G(\phi)$, i.e., 
\begin{align}
G_{\text{azim}} = \max\left( G(\phi)\right) - {<}G(\phi){>}, \ [\text{dB}].
\label{eqn:azimGain}
\end{align}
Accordingly, the effective azimuthal gain is obtained from \eqref{eqn:azimGain} by using the measured power {over azimuth $P(\phi)$, i.e.,  
\begin{align}
G_{\text{azim-effective}}= \max\left( P(\phi)\right) - {<}P(\phi){>}, \ [\text{dB}]. 
\label{eqn:Gain-eff}
\end{align}
}

{Note that the transmit antenna gain\footnote{{Nominal Tx antenna gain was used in path gain calculation since it was observed that the effective gain  of the  Tx antenna is very close to its nominal  due to its wide beamwidth (55$^\circ$) and the relatively small angular spread.}}   $G_{\text{T}}$ and the receive antenna elevation gain\footnote{{The low vertical angle spread\cite{TAP_6, TAP_7} means that the nominal elevation gain of the Rx antenna would be very close to its nominal value.}} $G_{\text{elev}}$ have been removed from the azimuthally averaged received power ${<}P(\phi){>}$. Therefore, the path gain obtained using \eqref{eqn:PL} is compatible with 3GPP models\cite{TAP_7} established for omni-directional Tx and Rx antennas.}     

 \section{Path gain measurements and models for suburban street canyon}\label{sec:pathGain}

\subsection{Outdoor-outdoor LOS with ground reflection}\label{sec:PL_tworay}

\begin{figure} 
	\centering
		\includegraphics[width=0.95\figwidth]{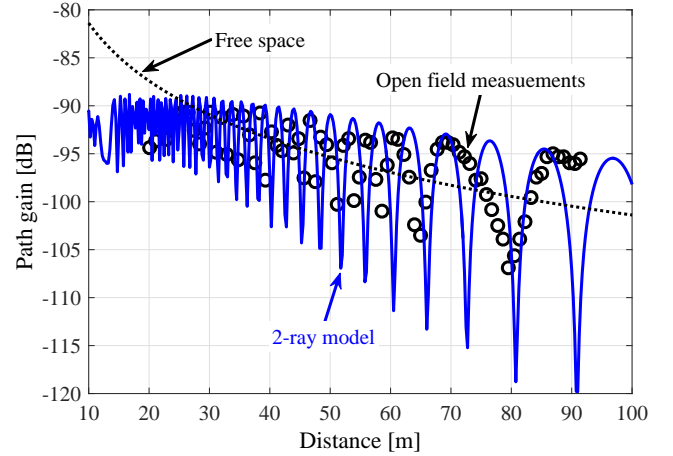}
	\caption{LOS path gain measurement (black circle) is in reasonable correspondence to the 2-ray model (solid line), where observed peak power amounts up to 6 dB above free space (dotted line). The 10$^\circ$ Rx horn was on van mast 3 m above ground and the 55$^\circ$ Tx horn 1 m above ground.}
	\label{fig:two-ray}
\end{figure}

An outdoor LOS calibration was carried out in an open field where the 10$^\circ$ Rx horn was on the van mast 3 m above ground and the 55$^\circ$ Tx horn was 1 m above ground. 
{For this measurement the antennas were aimed at each other, i.e. no azimuth rotation was used.} 
The measured path gain is shown in Fig.~\ref{fig:two-ray} together with a model from 2-ray theory which includes {the actual vertical misalignment of 2 m} and ground reflectivity, {which is computed as a reflection coefficient from a half-space with dielectric constant of 5 and loss tangent of 0.1}. The data exhibits constructive and destructive superposition in agreement with theory, with peaks about 6 dB above free space, and nulls that are more than 15 dB deep. Both observed and modeled path gains for distances shorter than 30 m are lower than free space caused by vertical misalignment, where links that are 20 m or shorter will be out of the 10$^\circ$ receive beam {in elevation plane}.

\subsection{Same-street outdoor-outdoor with vegetation}\label{sec:PL_samestreet}

\begin{figure} 
	\centering
		\includegraphics[width=\figwidth]{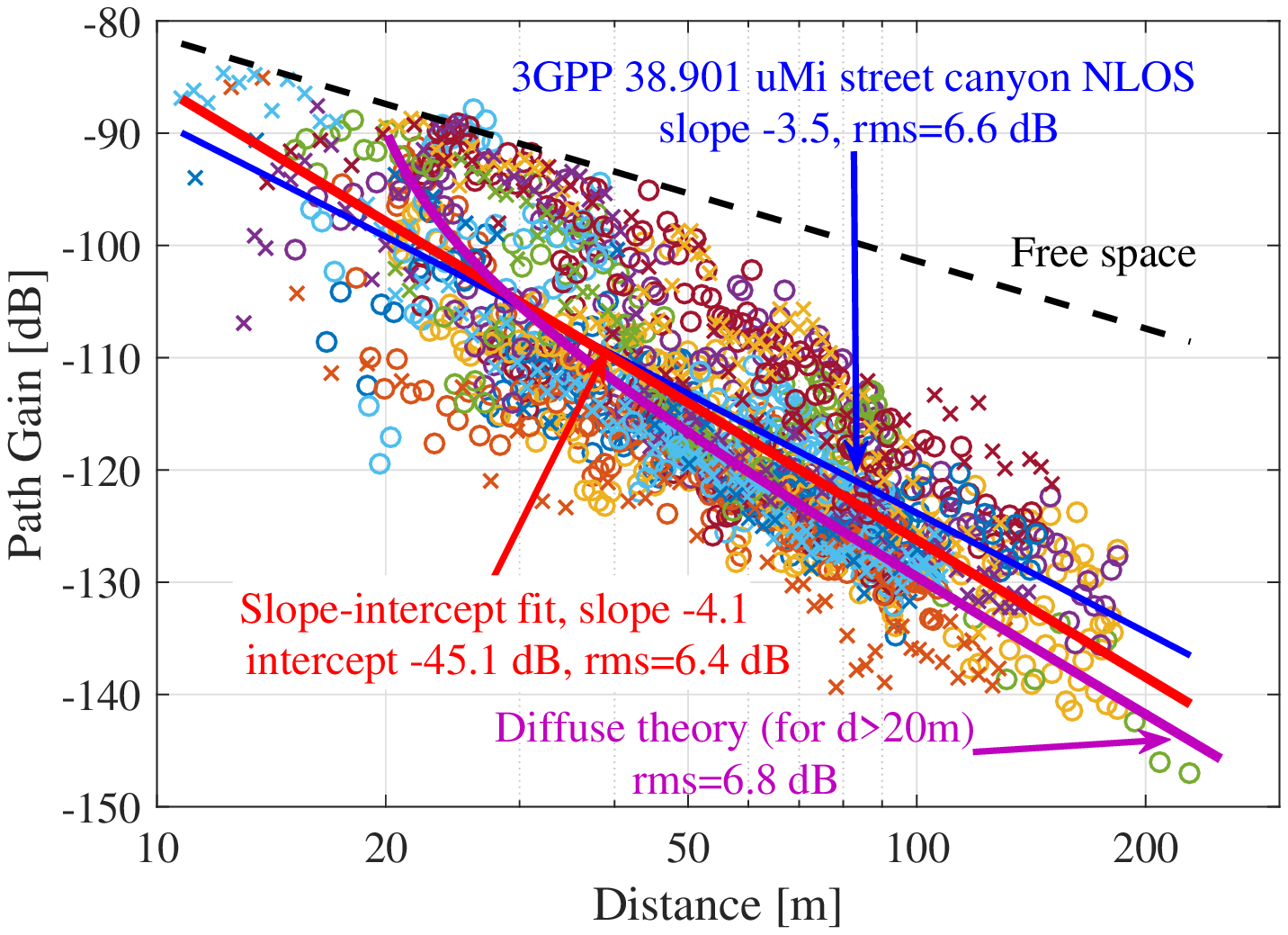}\\
		\includegraphics[width=\figwidth]{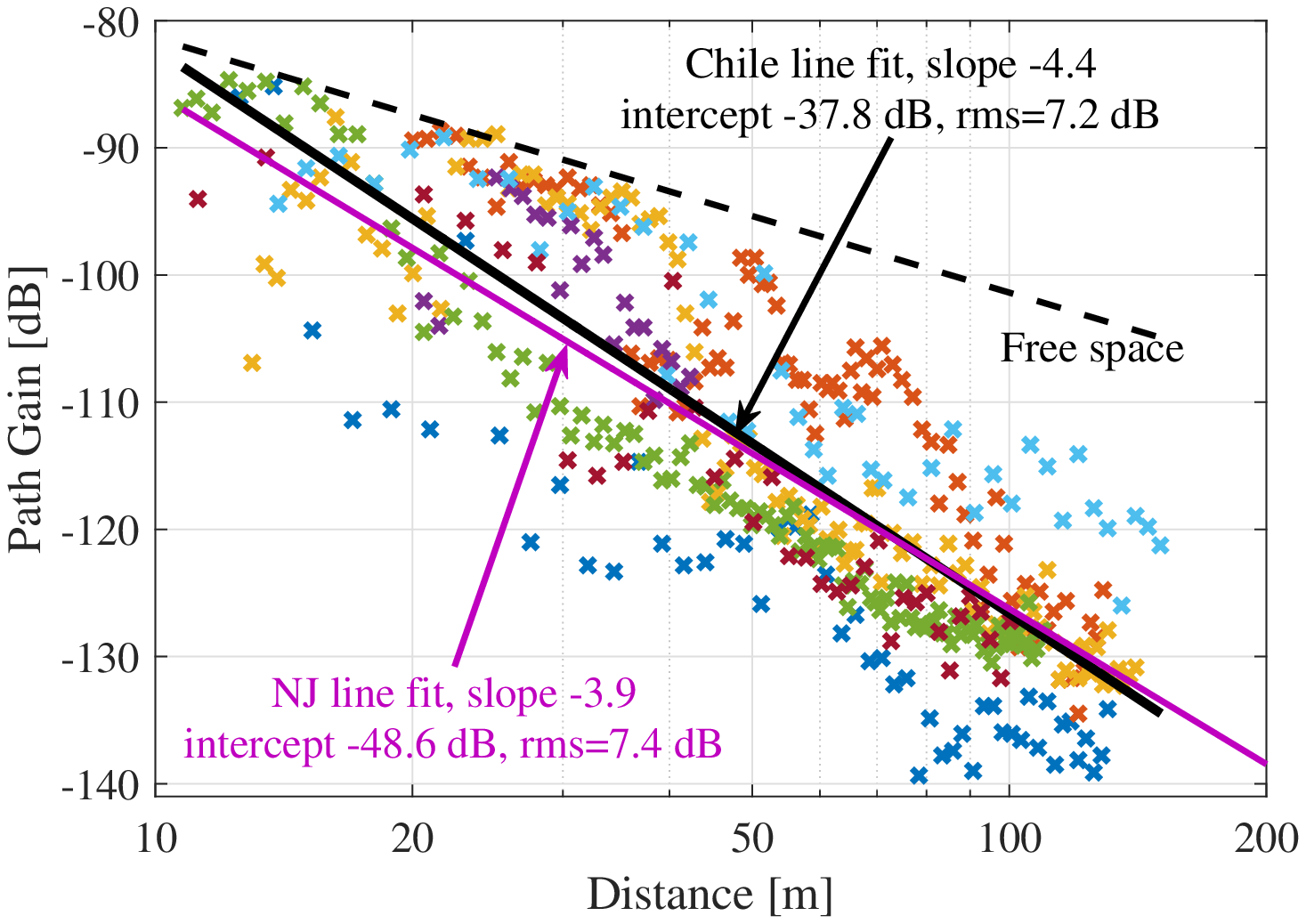}
	\caption{(Upper) Measured path gain for over 1700 links ranging up to 200 meters along each street (differs in color) in NJ (1322 links, marker o) and Chile (442 links, marker x) resulted in a slope of -4.06, intercept of -45.1 dB, and RMS deviation of 6.4 dB; (Lower) Comparing the Chile dataset against the NJ slope-intercept fit resulted in less than 0.3 dB increase of RMS error (7.2 dB RMS of Chile dataset against its own fit). }
	\label{fig:same-street}
\end{figure}

{Same-street outdoor-outdoor measurements were done with the rotary Rx moving along the street and the Tx placed near the exterior of a street-facing window. The Tx was aimed at a fixed angle along the street to illuminate the entire Rx route as previously described in Sec.~\ref{sec:scenario}.}
Path gain for each measured link is computed by averaging power over all angular directions to estimate the local average power as would be obtained from a spatial average of omni-antenna measurements. 
Measured same-street path gain as a function of distance $d$ [m], shown in Fig.~\ref{fig:same-street} (upper), was found to be well represented by the power law. The slope-intercept fit with  90\% confidence intervals~\cite{TAP_26} is given by
\begin{align}
 {	P_{\text{same-street}} = A + 10n\log_{10} (d) + \mN(0, \sigma), \ [\text{dB}], }\label{eqn:same_street}\\
 { A {=}-45.1{\pm}1.4,\ n {=} -4.06{\pm}0.08,\ \sigma{=}6.4, }\nonumber
\end{align} 
where $d$ is the Euclidean distance in meters between base and user terminal, $A$ is the 1-m intercept, $n$ is the slope of distance (in dB scale), and $\sigma$ is the RMS deviation representing shadow fading. As compared to free space, it has a 25 dB excess loss at 100 m. {Note that a significant data set facilitates reliable characterization of path gain models. With over 1700 links, the 90\% confidence intervals are very tight as shown in \eqref{eqn:same_street}.}

A model derived from diffuse theory \cite{TAP_21}, superimposed on Fig.~\ref{fig:same-street} (upper), had a slope of 4.0 and RMS error of 6.8 dB, comparable to the line-fit predictions, which is notable for a theory unadjusted to data and given the fact that the impact of vegetation blockage loss was here accounted for by modeling the layer of non-contiguous vegetation as a contiguous diffuse media. We note that the slope of 4.06 obtained from slope-intercept fit is very close to the slope predicted by the model derived from diffuse theory. The 3GPP 38.901 \cite{TAP_7} urban micro street-canyon NLOS model (indicated by the solid blue line in Fig.~\ref{fig:same-street}), which is intended to use for NLOS propagation in urban streets, had an error of 6.6 dB, comparable to the 6.4 dB RMS deviation from the linear fit to data. It should be noted, however, the 38.901 model prescribes the use of LOS and NLOS formulas, based on specified Probability of Line of Sight, that is set to 1 for ranges under 18 m and decreases to 0.5 at 50 m and 0.25 at 100 m. In LOS, the path gain is specified as being close to free space. Over 90\% of measurements in Fig.~\ref{fig:same-street} (upper) had excess losses relative to free space of over 10 dB, even at ranges of 50 m, despite being on the same street.

	Measured  {path gain datasets from NJ and from Chile have similar line fits and spread.} Comparing the Chile dataset against the NJ slope-intercept fit, as shown in Fig.~\ref{fig:same-street} (lower), resulted in less than 0.3 dB increase of the RMS error ( {as compared to the 7.2 dB RMS derivation from its own line fit}). Linear regression results of the measured path gain for the two datasets and their combination are summarized in Table~\ref{tab:same-street}. Applying the \emph{Common-Slope Cross-Comparison} method \cite{TAP_10} to the two datasets resulted in a common slope of 4.05 and a 1 dB gap between the two intercepts with combined RMS fitting error of 6.4 dB. Therefore, we can conclude that the empirical model presented in \eqref{eqn:same_street} is robust.

\begin{table}
	\centering
		\caption{Summary of same-street path gain measurements {and models with 90\% confidence interval\cite{TAP_26}.}}
	\label{tab:same-street}
	\begin{threeparttable}
		\begin{tabular}{|c|c|c|c|c|}
		\hline  
	&	Number of 	& \multicolumn{3}{|c|}{Path Gain}\\
	\cline{3-5}
&  links  & 		slope	&  intercept	&  RMS \\
\hline
NJ     &	1322 & -3.87 $\pm$ 0.09 & -48.6 $\pm$ 1.6 dB	& 6.0 dB\\
\hline
Chile	 & 442	 & -4.44 $\pm$ 0.16 & -37.8 $\pm$ 2.8 dB	& 7.2 dB\\
\hline
Combined& 1764 & -4.06 $\pm$ 0.08	& -45.1 $\pm$ 1.4 dB  &	6.4 dB\\
\hline	
		\end{tabular} 
	\end{threeparttable}
\end{table}

\begin{figure} 
	\centering
		\includegraphics[width=0.9\figwidth]{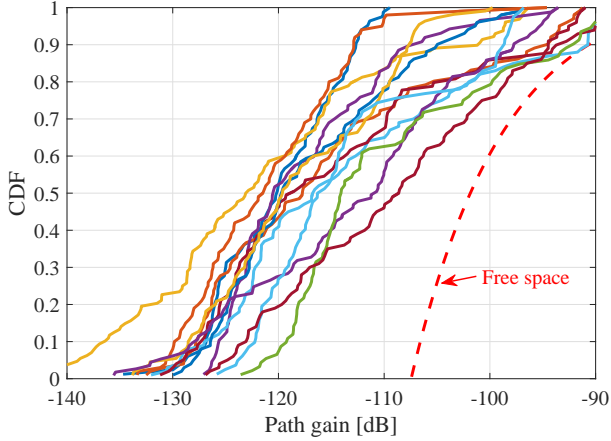}
	\caption{Distributions of measured path gain for each street separately: streets are quite different from each other, with medians spanning a range of 13 dB. Path gain for free space is plotted as reference (red dashed line).}
	\label{fig:CDF_path}
\end{figure}

 The measured path gain distributions for ranges of 20 m to 200 m are plotted separately for each street in Fig.~\ref{fig:CDF_path}. It may be observed that distributions vary substantially, with medians spanning a range of 13 dB. This illustrates the importance of collecting sufficient data over multiple streets and houses to get representative path loss. Modeling one street based on a data fit to data collected on another street may lead to some 13 dB (median) error in coverage.

\subsection{Other-street outdoor-outdoor measurement}\label{sec:PL_otherstreet}	

	To quantify coverage from lamppost-mounted base stations to terminals located in a different street {and to assess interference between neighboring streets}, we collected data for over 180 ``other-street'' links where the user terminal was placed at a house of one street, and the base moved along a parallel street separated by one block of 30 to 80 m width. {The rotary Rx was moving along the parallel street, gathering power measurements from azimuthal scans, and the Tx antenna was aimed at a fixed angle towards the Rx to illuminate its entire route.} 
	
\begin{figure} 
	\centering
		\includegraphics[width=0.98\figwidth]{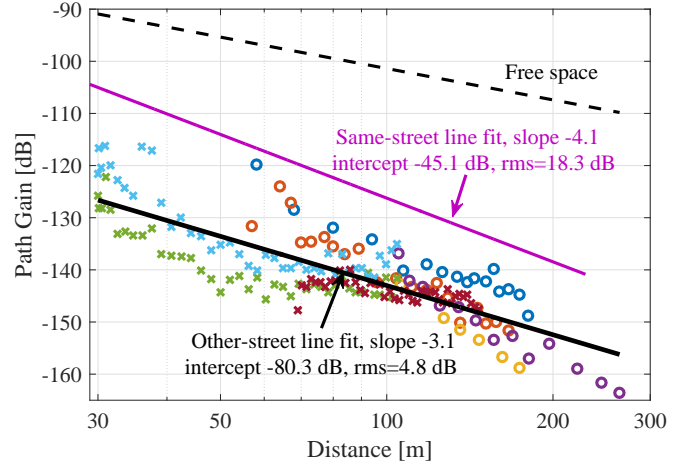}
	\caption{Measured path gain for over 180 other-street links along each street (differs in color) resulted in a slope of -3.13, intercept of -80.3 dB, and RMS error of 4.8 dB. As compared to free space, it has a 42 dB excess loss at 100 m. Using the same-street model of \eqref{eqn:same_street} to represent other-street data resulted in RMS error of 18.3 dB.}
	\label{fig:other_street}
\end{figure}

	Measured other-street path gain, shown in Fig.~\ref{fig:other_street}, was found to be well represented by
\begin{align}
P_{\text{other-street}} =  {A + 10n\log_{10} (d) + \mN(0, \sigma), \ [\text{dB}],} \label{eqn:other_street} \\
  {A {=}-80.3{\pm}4.1,\ n {=} -3.13{\pm}0.21,\ \sigma{=}4.8.} \nonumber
\end{align}
As compared to free space, it has a 42 dB excess loss at 100 m, in contrast to the 25 dB excess loss in the same-street model. Coverage from under-clutter base stations located in other streets is thus very limited, and the potential interference from other-street base stations is likewise mild, since the latter is typically 10 to 17 dB lower than the same-street signal even if the two links have the same distance. We note that using the same-street model \eqref{eqn:same_street}, to represent other-street data, resulted in RMS error of 18.3 dB, justifying the use of a separate model \eqref{eqn:other_street} for other-street links (for example, to assess interference or the viability of coverage from base stations on parallel streets).

\subsection{Outdoor-outdoor visual LOS links}\label{sec:PL_vLOS}

\begin{figure}  
		\centering 	
		\includegraphics[width=0.9\figwidth]{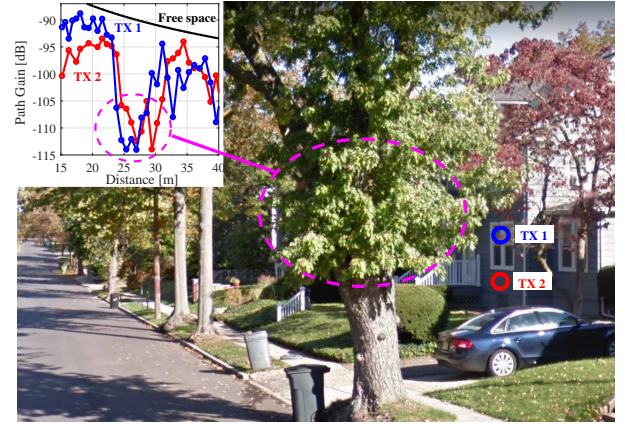} 
	\caption{Empirical observation of (partial) tree blockage from two different Tx antenna heights (TX1 at 3 m and TX2 at 1.5 m).  The shortest distance from the rotating Rx to the branches/leaves is about 2 m during the measurement.}
	\label{fig:tree_blockage}
\end{figure}

Scatterers/obstructions such as tree branches/leaves may partially block the {direct path between the Tx and Rx even for  same-street outdoor-outdoor measurements described in  Sec.~\ref{sec:PL_samestreet}}.
For links ranging from 30 m to 100 m, the maximum radius of the first Fresnel zone \cite{TAP_11} is several tens of centimeters at 28 GHz. This may leave ample opportunity for the direct path to survive, but  partial blockage within the first radio Fresnel zone can potentially cause significant power drop even in the presence of a visual LOS path.  We labeled these links that have a visual LOS path, determined by using a flashlight at the one end of the links and observed at the other end, as visual LOS. We plotted in Fig.~\ref{fig:tree_blockage} some empirical observations of the effect of (partial) blockage by tree branches. {As the Rx moves away from the Tx,} the link changes from LOS to NLOS and then back to LOS. Partial blockage was observed to result in 5 to 10 dB losses over free space propagation. In contrast under total blockage, more than 20 dB excess loss was observed.

	\begin{figure} 
	\centering
		\includegraphics[width=0.98\figwidth]{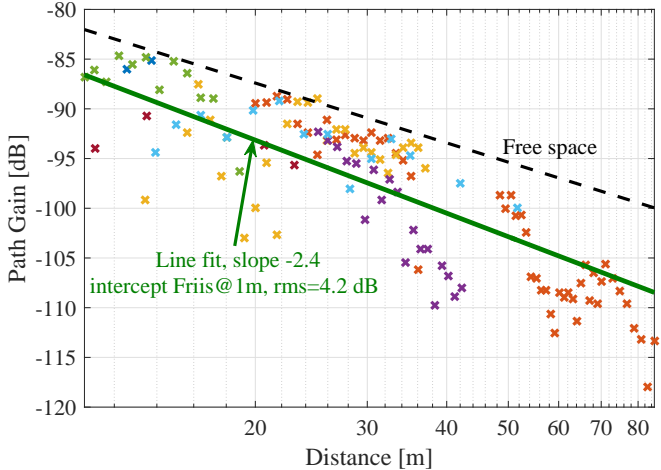}
	\caption{Measured path gain of visual LOS links (streets labeled by color). Obstruction within the first Fresnel zone significantly degrades the link quality.}
	\label{fig:vLOS}
\end{figure}

 {Characterizing the path gain of visual LOS links would allow the performance evaluation of the FWA deployment under what would be expected to be more} favorable condition. Data of visual LOS links are obtained from same-street outdoor-outdoor measurements described in  Sec.~\ref{sec:PL_samestreet} where the Tx  is placed either at 1 m or 3 m height on the street-facing exterior wall, and the base (Rx) is moving along the street to collect measurements. 
We plotted their path gain versus distance for these locations in Fig.~\ref{fig:vLOS}. Using linear regression with fixed intercept of Friis path gain at 1 m (i.e., -61.4 dB at 28 GHz), the visual LOS path gain dependence on distance $d$ was found to be well represented by
\begin{align}
P_{\text{visual-LOS}} &=  {-61.4 + 10n\log_{10} (d) + \mN(0, \sigma),  \ [\text{dB}],} \label{eqn:vLOS}\\
&  {n{=}-2.44{\pm}0.21,\ \sigma{=}4.2. }\nonumber
\end{align}
Here free space 1-m intercept is adopted because our dataset is now limited to the points closest to that condition. Optimization of both the intercept and slope resulted in no more than a 0.2 dB reduction in the RMS error.

      As compared to free space, the measured excess loss can be up to 15 dB for links at various distances as shown in Fig.~\ref{fig:vLOS}.  The visual LOS condition thus does not imply free space propagation loss.

\section{Azimuth gain measurements}\label{sec:azim}

Nominal antenna gain can be degraded as a result of significant angular spread compared to beamwidth in scattering environments \cite{TAP_14}, and a large gain reduction can significantly degrade the performance of mmWave systems~\cite{TAP_22}. To reliably quantify this effect, we measured the power as a function of azimuth as the Rx antenna was spinning. For every Tx-Rx pair the effective azimuth gain was computed as the ratio of the maximum power to the average over all directions {using \eqref{eqn:Gain-eff}}, as justified in \cite{TAP_8} under the assumption of uncorrelated scattering. This allows us to directly quantify the effective beamforming (BF) gain, {in contrast to the indirect approach} adopted by \cite{TAP_14} where an estimated channel {power angular distribution was used to deduce the BF gain reduction}. The resulting distribution of azimuth gains is shown in Fig. 9. The azimuth gain exceeded in 90\% of cases at the 3 m base station in the street is found to be 10.2 dBi, as opposed to the nominal gain of 14.5 dBi in azimuth, i.e., a 4.3 dB gain reduction. 

\begin{figure} 
	\centering
		\includegraphics[width=0.9\figwidth]{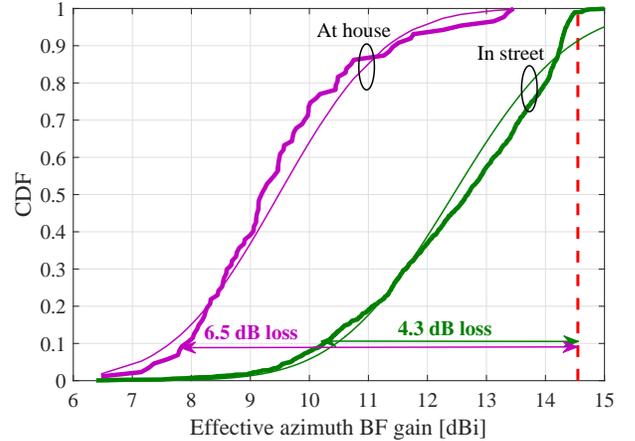}
	\caption{CDFs of effective azimuth gain of the rotary horn antenna, where the degradation is 4.3 dB at $10^{\text{th}}$ percentile when placed in streets (as base station) and 6.5 dB at houses (as CPE). Log-normal fits (thin lines) have mean of 12.4 dB and 9.5 dB, respectively, with a standard deviation of 1.5 dB.}
	\label{fig:CDF_azim}
\end{figure}

To investigate the possible benefits of using a more directive antenna at the customer’s premises, we placed the rotary horn near the house wall (in proximity of vegetation), and obtained an azimuth gain of 8 dBi at the 10th percentile, a 6.5 dB gain reduction 
(leftmost curve in Fig.~\ref{fig:CDF_azim}). The observed effective azimuth gains replace the nominal antenna gain in link budget calculations. 

We found that the effective gain distribution for the base station varies mildly from street to street, with medians ranging from 11.2 to 14 dBi, and no clear distinction has been observed between same-street and other-street measurements. Log-normal fits (thin lines) to the empirical CDFs of effective azimuth gain ``in street'' and ``at the house'' have means of 12.4 dB and 9.5 dB, respectively, with standard deviation of 1.5 dB. Using a more advanced base antenna of the same aperture (e.g. combining multiple beams), which requires channel   information and additional RF chains, may bring additional gain, upper-bounded by 4.3 dB in 90\% of locations in this environment. 

\section{Potential gain of beam re-aim in azimuth}\label{sec:switch}

Temporal variation  of received power along azimuth directions, mainly caused by wind-blown branches/leaves, could potentially degrade the quality of service for FWA users even if neither end moves.  To quantify the potential gain of changing beam directions to compensate for such temporal fades, we have extracted at each link the power samples along a direction that provides the best-on-average power over time, referred as the best (a posteriori) angle. For each link, this results in no less than 37 power measurements in a fixed direction over a time window of 10 seconds.

Such temporal fades along a fixed direction have been found  to be well represented by Rician distributions for  FWA links in suburban~\cite{TAP_24} and in open spaces with tree blockage~\cite{TAP_25} at lower frequency bands (1.8 to 1.9 GHz). Temporal fades measured in a street canyon at 60 GHz~\cite{VC97}  also reported good agreement in CDF with the Rician model, but the agreement in level crossing rate is less satisfactory for data collected using a moving terminal. On the other hand, sparse multipath mobile channels, as may occur at mmWave frequencies, have exhibited non-Rician behavior~\cite{PR13,UWB_fading}, and small-scale spatial fading may be better represented by the two-wave with diffuse power (TWDP) fading model\footnote{TWDF and its extensions, such as~\cite{TWDP_19, FTR_17}, degenerate to Rician when the number of specular rays is reduced from two to one.} \cite{TWDP_19, FTR_17}.  

\begin{figure} 
	\centering
		\includegraphics[width=0.9\figwidth]{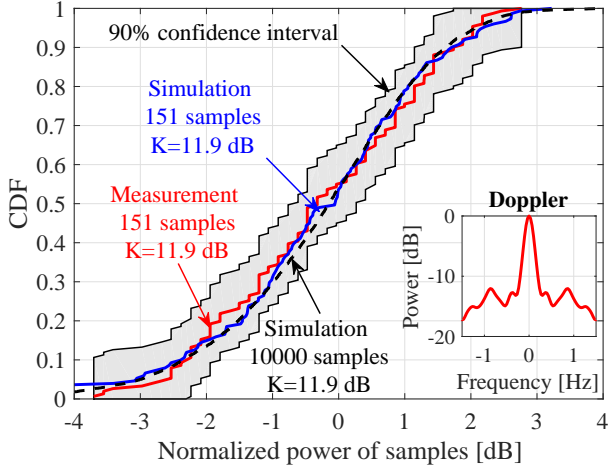}
	\caption{ {CDF of temporal power variation along the best angle for a FWA link measured in NJ. Insert is the estimated Doppler spectrum obtained following the method developed in~\cite{DGMS03}. CDFs of simulated Rician samples with the same $K$-factor  are also plotted for reference. All three CDFs match well with one another and the variations are small as compared to the $90\%$ confidence interval\cite{TAP_27} of the measured data.}}
	\label{fig:CDF_Rician_fit}
\end{figure}

We tested the distribution of temporal fades measured at 28 GHz for a FWA link and compared its CDF to those obtained using simulated samples from a Rician distribution with the same $K$-factor as estimated using the method of moments (MoM) \cite{TAP_12, TAP_24, TAP_25}. As shown in Fig.~\ref{fig:CDF_Rician_fit},  the temporal fades along the best (a posteriori) angle were well represented by a Rician distribution.
We also computed the Doppler spectrum of the link, shown as an insert of Fig.~\ref{fig:CDF_Rician_fit}, following the method developed in~\cite{DGMS03}. The Doppler spectrum was found to be peaked at 0 Hz with exponential decay, and the spectral density falls by 13 dB between 0 and 0.3 Hz. This is consistent with the theoretical model~\cite{TPE02} developed for moving scatterers and the empirical observation of Rician temporal fades induced by wind-blown leaves~\cite{DGMS03} for fixed wireless links at 5.3 GHz.

\begin{figure} 
	\centering
		\includegraphics[width=0.9\figwidth]{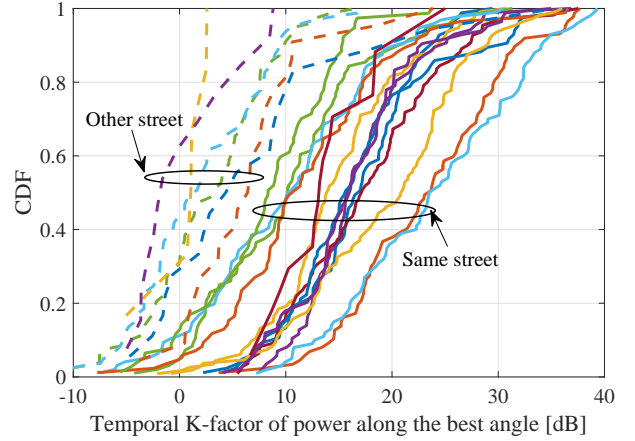}
	\caption{CDFs of temporal $K$-factor along the best (a posteriori) angle for each link measured in NJ, with median values range from -2 to 24 dB. Solid lines are same-street links (median 16 dB) and dashed lines are for other-street (median 2.5 dB). The ensemble of temporal $K$-factor for all the same-street links is well represented by a log-normal distribution with mean 16.7 dB and standard deviation 8.9 dB.}
	\label{fig:CDF_KdB}
\end{figure}

Given the observed Rician behavior, we computed the  temporal $K$-factors of the observed signal envelopes for each street, as shown in Fig.~\ref{fig:CDF_KdB}, where the median value of each street ranges from -2 to 24 dB. The ensemble of temporal $K$-factors for all the same-street links is well represented by a log-normal distribution,  {consistent with the findings reported in \cite{TAP_24, TAP_25},} with mean 16.7 dB and standard deviation 8.9 dB. The same-street links in general have higher $K$-factors (median 16 dB) than other-street links (median 2.5 dB). This can be explained by the fact that there are more  vegetation and houses  between the Tx and Rx for other-street links. Moving vehicles may also contribute to the lower $K$-factors observed for other-street links: there may have been moving vehicles in the streets where the CPE was located when the base station was collecting power measurements in another street without traffic.  In contrast, same-street measurements were collected during intervals where no moving vehicles were observed.

	\begin{figure} 
	\centering
		\includegraphics[width=0.9\figwidth]{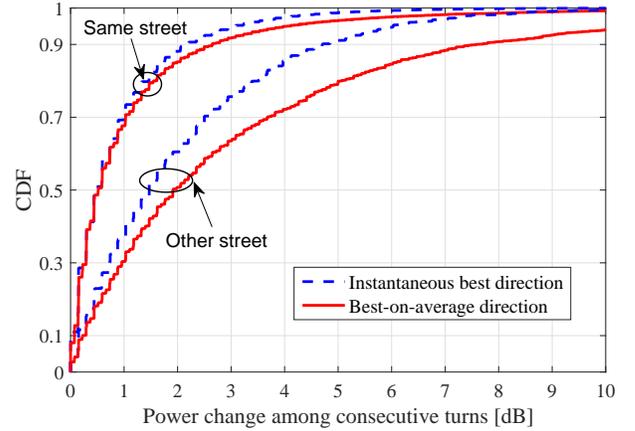}
	\caption{CDFs of power change among consecutive turns for the instantaneous best direction of each turn (blue dashed lines) and for power along the best-on-average direction (red lines) for each link. For more than 90\% of all same-street links, power fluctuation is less than 3 dB.
	}
	\label{fig:CDF_fluctuation}
\end{figure}

         For the angular spectrum measured at each link, we choose the highest power direction for each turn, referred as instantaneous best direction. This creates a sequence of instantaneous best directions over time, relevant when the azimuthal scan measurement is fast enough to assure best-aim is maintained over time. In Fig.~\ref{fig:CDF_fluctuation} we plot the CDFs of power change between consecutive turns for the instantaneous best direction of each turn and for power along the best-on-average (a posterior) direction for each link. As seen, for same-street links the fluctuation of the best per-rotation power sample and the sample for the best-on-average angle is very small. This suggest that beamswitching will offer very modest gains.  More than 90\% of all same-street links experienced less than 3 dB power fluctuation between consecutive turns (about 200 ms) and the corresponding beamswitching gain is less than 1 dB. However, power fluctuation is more severe for other-street links, close to 8 dB at the 90th percentile along the best angle, with a potential beamswitching gain of 3 dB.

\section{Estimates of achievable outdoor-to-outdoor downlink rates for 90\% coverage}\label{sec:Rate}

To evaluate the edge rate for 90\% coverage based on the findings reported above, we computed the Shannon rate for outdoor CPEs {that will be available to 90\% of users at a given range. Equivalently, we determined the SNR exceeded for 90\% of CPEs at this range, denoted as $\gamma_{90\%}(d)$, and from it the corresponding Shannon rate
\begin{align}
   R_{90\%}(d) = W\log_2(1+\gamma_{90\%}(d)),   \label{eqn:Shannon}
\end{align}
where $W$ is the bandwidth. }

The lamppost mounted base station is assumed to have 51 dBm EIRP and bandwidth of 800 MHz\footnote{Maximum of 850 MHz bandwidth at 28 GHz band was allocated by FCC. When channel dynamics is significant (e.g., by moving vehicles), which is not studied here, using all the available bandwidth might be counter-productive, see \cite{TAP_13}, since the channel estimation cost and penalty might be overwhelming to maintain coherent transmission.}, and the outdoor CPE has 9 dB noise figure and 11 dBi antenna gain. A cellular small cell operating at 2 GHz band (using 30 dBm transmit power and 5 dBi antenna gain for both the base station and the UE) {is presented as a baseline for comparison. Key parameters used for rate calculation for the 28 GHz system as well as the 2 GHz system are listed in Table~\ref{tab:simulation}.}

Both transmitter and receiver are here assumed to be equipped with a single RF chain and use simple directional antennas, pointed adaptively so as to get maximum power. Given the modest degradation of azimuth gain and modest temporal variations found in this campaign, such a simple system is seen here as a reasonable reference. For each link, the path gain was generated from a random process as described by either \eqref{eqn:same_street} or \eqref{eqn:other_street} for same-street and other-street scenarios, respectively, and the effective azimuth gain was generated using the log-normal fits from Sec.~\ref{sec:azim}. The path loss and angular spread for the 2 GHz system are generated using 3GPP 36.814 UMi NLOS models specified in \cite{TAP_9}. Channel dynamics caused by wind-blown leaves/branches was not included in the calculation as they were found to be a minor effect in our measurement. For each base-CPE separation distance, 10000 links are generated randomly and independently, from which we obtain the Shannon rate with 90\% coverage guarantee for that distance {using \eqref{eqn:Shannon}}.

\begin{table}
	\centering
		\caption{Parameters for 90\%-coverage downlink rate calculation.}
	\label{tab:simulation}
	\begin{threeparttable}
		\begin{tabular}{|c|c||c|}
		\hline  
 Carrier frequency  &  28 GHz	&  2 GHz \\
\hline
 Signal bandwidth  &   800 MHz & 20 MHz\\
\hline
Lamppost BS  & 28 dBm, 23 dBi & 30 dBm, 5 dBi\\
\hline
Outdoor CPE & 11 dBi &  5 dBi\\
\hline	
Noise figure & 9 dB  & 9 dB\\
\hline
Path loss & \eqref{eqn:same_street} or \eqref{eqn:other_street} & 3GPP 36.814 \cite{TAP_9}\\
\cline{1-2}
Gain reduction & results of Sec.~\ref{sec:azim}  & UMi NLOS \\
\hline
		\end{tabular} 
	\end{threeparttable}
\end{table}

	\begin{figure} 
	\centering
		\includegraphics[width=0.9\figwidth]{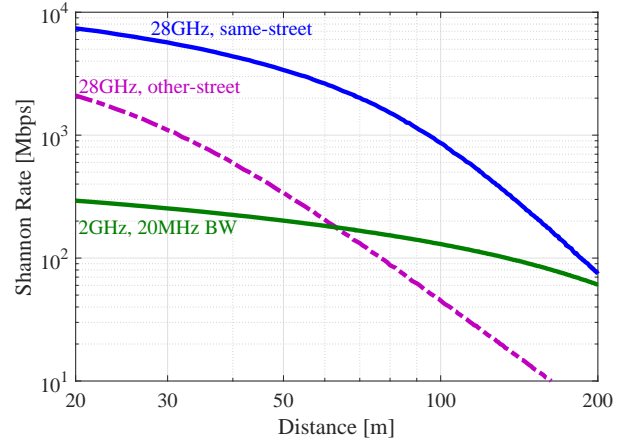}
	\caption{90\%-coverage downlink edge rate to outdoor CPE where the base station has 28 dBm transmit power and 23 dBi antenna gain and the CPE has 11 dBi antenna gain with 9 dB noise figure. Path loss and effective azimuth gain are generated using measured path loss models and azimuth gain distributions, and the rate is Shannon rate with 800 MHz bandwidth. The edge rate will be significantly reduced when moving CPE to indoor. As a reference, we also plot the rate for a cellular small cell system operating at 2 GHz (30 dBm transmit power, 5 dBi gain antennas for base station and UE) using bandwidth of 20 MHz (solid green line). }
	\label{fig:Rate}
\end{figure}

The resulting rate-distance plots are shown in Fig.~\ref{fig:Rate}. It was found that 1 Gbps downlink rate can be delivered to outdoor CPE in the same street for distance up to 100 m, and the rate decreases to 80 Mbps when the distance increases to 200 m. If the base station and the CPE are in different streets, 100 Mbps downlink rate can be delivered up to 80 m with 90\% coverage guarantee. Since outdoor-to-indoor penetration loss is large, 9 to 17 dB median loss as reported in \cite{TAP_10} for NJ suburban residential homes, the corresponding ranges will decrease by a factor of 2 or more. Compared to a cellular small cell operating at 2 GHz band, 28 GHz cell can deliver a higher rate (with 90\% coverage guarantee) than the 2 GHz system over the 200 m range for same-street CPEs. Should 100MHz bandwidth be available for the 2 GHz system (for example, via carrier aggregation), the rate at 100 m with 90\% coverage guarantee would be increased to 400 Mbps, about half of what can be supported by the 28 GHz system using 800 MHz.  On the other hand, a conventional 2 GHz system using 20 MHz would deliver higher rates than an 800 MHz-wide 28 GHz system beyond 60 m in general NLOS conditions, where the base station is not on the same street as the terminal being served.

\section{Conclusions}\label{sec:conclusion}

 Extensive measurements (over 2000 links, each containing at least 37 full azimuthal scans with power measurements over 360 degree) of path loss, achievable azimuth gain and temporal fading were collected and characterized statistically at 28 GHz in suburban environments in NJ and Viña del Mar, Chile. The same-street path loss fit has a slope of 4.1 with 25 dB excess loss over free space at 100 m, and the other-street path loss fit has a slope of 3.1 with 42 dB excess loss at 100 m. Links that have visual LOS path (assured using a flashlight) suffer up to 15 dB excess loss over free space over a wide range of distances. An ordinary 24 dBi directional antenna with adaptive pointing was found to lose up to 4.3 dB of gain in 90\% of suburban links. Using multiple chains within the same antenna aperture can help recover some of this loss, upper bounded by 4.3 dB. In the absence of moving vehicles, the benefit of rapidly re-aiming the beam at each azimuth scan is less than 2 dB. It was found that, with typical power and antenna gain configuration for FWA base station and CPE, 1 Gbps downlink rate can be delivered to an outdoor mounted CPE for up to 100 m from a base station deployed in the same street with 90\% coverage guarantee. If the base station and the CPE are in different streets, 100 Mbps downlink rate can be delivered up to 80 m with 90\% coverage guarantee. The corresponding downlink edge rate will be significantly reduced when moving outdoor CPE to indoor, highlighting the challenge for FWA deployments.

\section*{Acknowledgment}
The authors wish to acknowledge support by CONICYT under Grant Proyecto Basal FB0821, Fondecyt Iniciaci\'on 11171159, and Proyecto VRIEA-PUCV 039.462/2017 for supporting Mauricio Rodr\'iguez and Guillermo Castro. Many thanks to Hector Carrasco, Leonardo Guerrero and Rene Pozo for designing and building the platform, Cuong Tran for essential diagnostics and repair, and Alicia Musa for data collection software improvement.

\end{document}